\documentclass[aps,prb,letterpaper,superscriptaddress,10pt,twocolumn,floatfix,showkeys]{revtex4-2}

\usepackage[utf8]{inputenc}
\usepackage[T1]{fontenc}
\usepackage{amssymb}
\usepackage{amsmath}

\usepackage{graphicx}% Include figure files
\usepackage{dcolumn}% Align table columns on decimal point
\usepackage{comment}
\begin{document}

%\preprint{APS/123-QED}

\title{Spin-Phonon Relaxation of Boron-Vacancy Centers in Two-Dimensional Boron Nitride Polytypes}% Force line breaks with \\

\author{Nasrin Estaji}
\affiliation{Department of Physics, Isfahan University of Technology, Isfahan 84156-83111, Iran}

\author{Ismaeil Abdolhosseini Sarsari}\email{Contact author: abdolhosseini@iut.ac.ir}
\affiliation{Department of Physics, Isfahan University of Technology, Isfahan 84156-83111, Iran}

\author{Gerg\H{o} Thiering}
\affiliation{HUN-REN Wigner Research Centre for Physics, P.O.\ Box 49, H-1525 Budapest, Hungary} 

\author{Adam Gali}\email{Contact author: gali.adam@wigner.hun-ren.hu}
\affiliation{HUN-REN Wigner Research Centre for Physics, P.O.\ Box 49, H-1525 Budapest, Hungary}
\affiliation{Department of Atomic Physics, Institute of Physics, Budapest University of Technology and Economics, M\H{u}egyetem rakpart 3., H-1111 Budapest, Hungary}
\affiliation{MTA–WFK Lend\"{u}let "Momentum" Semiconductor Nanostructures Research Group, P.O.\ Box 49, H-1525 Budapest, Hungary}

\date{\today}% It is always \today, today,

\begin{abstract}
Two-dimensional (2D) materials hosting color centers and spin defects are emerging as key platforms for quantum technologies. However, the impact of reduced dimensionality on the spin-lattice relaxation time ($T_1$) of embedded defect spins—critical for quantum applications—remains largely unexplored. In this study, we present a systematic first-principles investigation of the negatively charged boron-vacancy (V$_{\text{B}}^-$) defect in monolayer boron nitride (BN), as well as in AA$^\prime$-stacked hexagonal BN (hBN) and ABC-stacked rhombohedral BN (rBN).
Our results reveal that the $T_1$ times of V$_{\text{B}}^-$ in monolayer BN and hBN are nearly identical at room temperature. Surprisingly, despite the symmetry reduction in rBN opening additional spin relaxation channels, V$_{\text{B}}^-$ exhibits a longer $T_1$ compared to hBN. We attribute this effect to the stiffer out-of-plane phonon modes in rBN, which activate spin-phonon relaxation at reduced strength. These findings suggest that V$_{\text{B}}^-$ in rBN offers enhanced spin coherence properties, making it a promising candidate for quantum technology applications.
\end{abstract}

\maketitle

%\tableofcontents

%\section*{Introduction}
\textit{Introduction.}
Room-temperature defect spins functioning as qubits are highly desirable for a range of quantum technology applications, particularly for probing biochemical processes. However, electron spin coherence is fundamentally limited by the spin-lattice relaxation time ($T_1$), which decreases rapidly with increasing temperature. As a result, only a few defect spins have been observed with relatively long $T_1$ lifetimes ($\gtrsim 10$$\mu$s) at room temperature~\cite{cambria2021state, cambria2023temperature, lin2021temperature, mondal2023spin, liu2025temperature, gale2025quantum}. Notably, these defect spins are found in materials composed of light elements from the second row of the periodic table (e.g., carbon in diamond, boron and nitrogen in boron nitride) or a combination of second- and third-row elements (e.g., carbon and silicon in silicon carbide). Phenomenological theories suggest that materials with high Debye frequencies are more likely to host qubits with long $T_1$ at room temperature~\cite{wolfowicz2021quantum}, consistent with these observations.

In this context, two-dimensional (2D) materials offer additional advantages. The spin-lattice relaxation rate ($1/T_1$) is expected to scale more slowly in 2D materials than in their three-dimensional counterparts, due to the reduced density of acoustic phonons contributing to spin relaxation~\cite{norambuena2018spin}. Combined with the promise of scalable qubit integration and proximity control, this makes 2D materials embedding defect spins an appealing platform for various quantum technology applications. However, truly freestanding 2D materials are difficult to realize experimentally, and interlayer interactions may influence the observed $T_1$ of defect spins embedded in 2D hosts.

\begin{figure}[h]
%    \centering
  \includegraphics[width=0.45\textwidth]{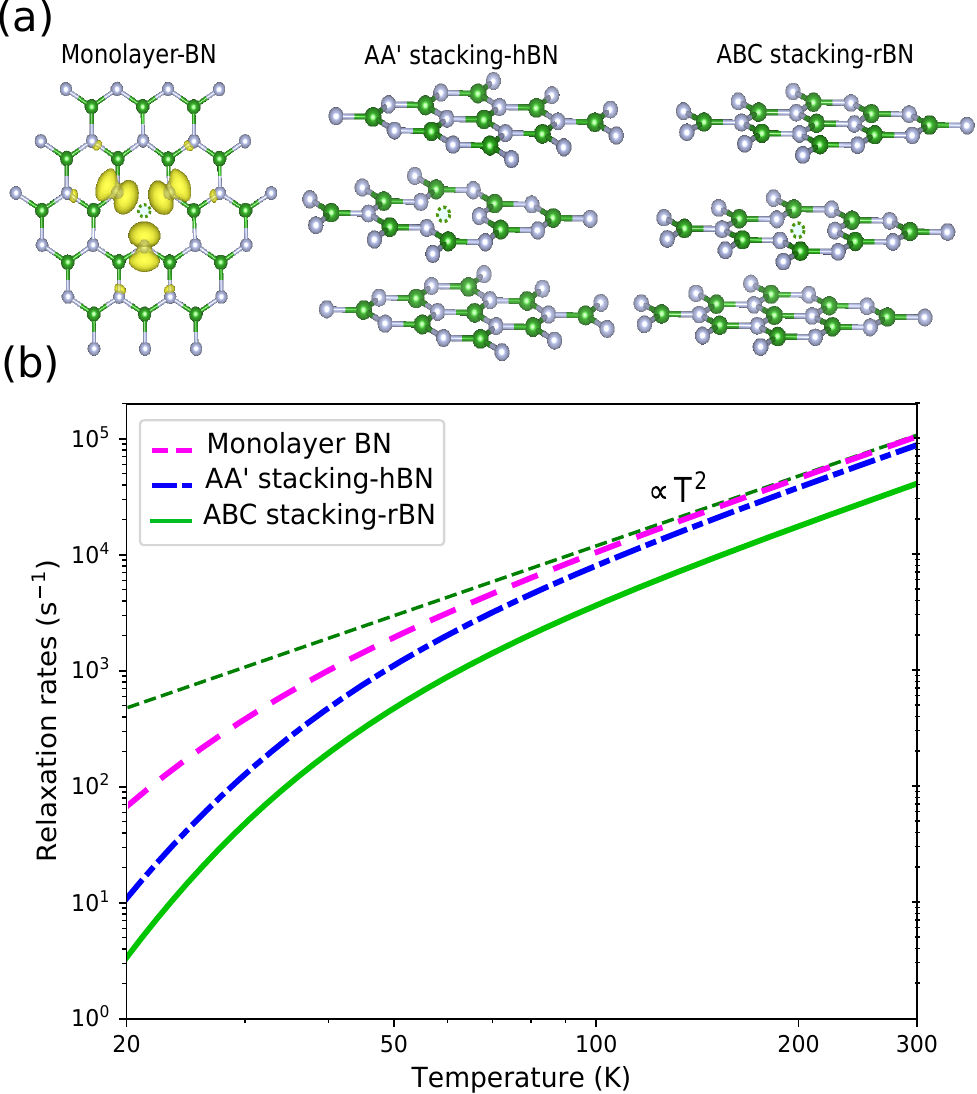}
    \caption{\label{fig:structs_rates} 2D BN material and computed spin-phonon relaxation rates of the embedded V$_{\text{B}}^{-}$ defect spin. (a) The atomic structure of freestanding monolayer BN, the AA$^\prime$-stacked hBN, where two adjacent BN layers are rotated by angle of $\pi$, and the ABC-stacked rBN, where the layers are successively shifted in the same direction by the interatomic distance. The vacant boron site is depicted as a semitransparent green ball where the spin density is localized on the three neighbor nitrogen atoms. (b) Computed two-phonon-assisted spin-phonon relaxation rates for monolayer BN, hBN and rBN. They show universal $T^2$ power-law slope at elevated temperatures with distinct shifts.}
\end{figure}

To investigate this issue, we selected the negatively charged boron vacancy (V$_{\text{B}}^-$) in the honeycomb lattice of boron nitride (BN), a prominent defect spin. Several factors motivated our choice:
(i) V$_{\text{B}}^-$ has a spin-triplet ground state ($S = 1$) and exhibits optically detected magnetic resonance (ODMR) in AA$^\prime$-stacked hexagonal BN (hBN), enabling coherent control at room temperature~\cite{ye2019spin, gottscholl2020initialization, gottscholl2021room, haykal2022decoherence, meng2022coherent, gong2023coherent, robertson2023detection}, with well-documented temperature-dependent $T_1$ data~\cite{gottscholl2021room, liu2025temperature};
(ii) theoretical $T_1$ values have been reported for V$_{\text{B}}^-$ in monolayer BN models~\cite{mondal2023spin};
(iii) in addition to hBN, the other stable BN polytype---ABC-stacked rhombohedral BN (rBN), shown in Fig.~\ref{fig:structs_rates}(a)—has been reported to host defect emitters~\cite{zanfrognini2023distinguishing, iwanski2024revealing} and defect spins~\cite{gale2025quantum}, likely originating from V$_{\text{B}}^-$.

In this study, we compute the spin-lattice relaxation rate of V$_{\text{B}}^-$ in monolayer BN, hBN, and rBN from first principles, using a theoretical framework previously validated for the nitrogen-vacancy center in diamond~\cite{cambria2023temperature}. For hBN, our results show good agreement with experimental data at elevated temperatures. We find that log-log $1/T_1$ exhibits an universal $\propto T^2$ dependence across all BN materials studied. Remarkably, our calculations reveal that V$_{\text{B}}^-$ in rBN has a longer $T_1$ than in hBN as plotted in Fig.~\ref{fig:structs_rates}(b). This is an unexpected outcome, as symmetry arguments would predict the opposite trend. Our result underscores the importance of first-principles approaches in accurately predicting spin-lattice relaxation times and positions V$_{\text{B}}^-$ in rBN as a promising defect spin candidate in 2D materials.
%

%\subsection{\label{sec:citeref}Spin-phonon relaxation rate}

%\section*{Methods}\label{sec:methods}
\textit{Methods.}\\
(a) \textit{Electronic structure and phonon calculations.}
%\subsection*{Electronic structure and phonon calculations}
The electronic structure calculation employs the supercell plane-wave projector-augmented-wave (PAW) method as implemented in the Vienna ab-initio simulation package (VASP)~\cite{kresse1996efficient}. Spinpolarized calculations are implemented using the generalized gradient approximation (GGA) with the Perdew-Burke-Ernzerhof (PBE) functional for electron exchange-correlation \cite{perdew1996generalized} in the calculation of atomic relaxation, phonons, and zero-field splitting (ZFS) tensors of V$_{\text{B}}^-$. We modeled the defect within supercell formalism. We applied 20~\AA\ vacuum size for the freestanding monolayer BN to separate the layers considered as isolated. For bulk hBN and rBN, the optimized interlayer distance is 3.3~\AA\ with the Grimme DFT-D3 method for dispersion correction. $\Gamma$-point sampling of the Brillouin-zone is used in all calculations. We employed a cut-off energy of 450~eV for the expansion of plane waves for the pseudo wavefunctions. The energy of the electronic iterations converged to $10^{-8}$~eV, and the force on the atoms converged to $10^{-4}$~eV/\AA\ in the minimum global energy of the adiabatic potential energy surface.  The phonons and associated normal coordinates are calculated by building up the Hessian matrix as the first numerical derivative of the forces acting on the atoms (see other details in Appendix B).

(b) \textit{Calculation of the spin-phonon relaxation rates.}
%\subsection*{Calculation of the spin-phonon relaxation rates}
The ZFS is originated from the dipolar electron spin-spin interaction in the considered systems, labeled as $D$-tensor. The $D$-tensor is calculated within the PAW-method~\cite{bodrog2013spin} as implemented by Martijn Marsman (see Appendix A for further details).

In this system, the spin-phonon interaction of the electronic ground-state $V_{\text{B}}^{-}$ with $S=1$ ground state is described by
\begin{equation}
\hat{V}=\overleftarrow{S}\overleftrightarrow{D}\overrightarrow{S}
\end{equation}
where $\overleftarrow{S}=\overrightarrow{S}{}^{\dagger}=(\hat{S}_{x}\;\hat{S}_{y}\;\hat{S}_{z})$, and  $\overleftrightarrow{D}$ is the $D$-tensor. 
The spin-phonon matrix elements are obtained by exploiting the dependence of the $D$-tensor on the normal coordinates ($R_i$) which is expressed as
\begin{align}
\label{eq:D-expansion}
    \overleftrightarrow{D}(R) &= \overleftrightarrow{D}(R=0)+\sum_{i}\frac{\partial{\overleftrightarrow{D}}}{\partial{R_i}}\Big|_{R=0}R_i+  \notag \\
    &+\frac{1}{2}\sum_{ij}\frac{\partial{\overleftrightarrow{D}}}{\partial{R_i}\partial{R_j}}\Big|_{R=0}{R_i}{R_j}
\end{align}
with a homebuilt script, which was originally developed to study nitrogen-vacancy centers in diamond, is used to extract the coefficients \cite{cambria2023temperature}. 
Dimensionless coordinates are expanded in terms of the phonon creation and annihilation operators $\hat{R_i}=({b_i}^\dagger + b_i)/{\sqrt{2}}$. The dominant contribution comes from the second-order derivatives~\cite{mondal2023spin, cambria2023temperature}. To evaluate the second-order derivatives, only diagonal terms are considered.  

The relaxation rate as a function of temperature ($T$) between initial and final spin states
$|m_s\rangle$ and $|m_s' \rangle$ by using the Fermi's golden rule in continuum limit describes as
\begin{equation}
\label{eq:gamma_msms}
\begin{split} & \Gamma_{(m_{s}m_{s}^{\prime})}(T)=\\
 & \frac{4\pi}{\hbar}\int_{0}^{\infty}d(\hbar\omega)n(\omega)[n(\omega)+1]F_{m_{s}m_{s}^{\prime}}^{(2)}(\hbar\omega,\hbar\omega)\text{,}
\end{split}
\end{equation}
where the spectral function $F_{{m_s}{m_s'}}^{(2)}(\hbar \omega, \hbar \omega)$ accounts for the phonon density of states and the spin-phonon coupling strengths (see Ref.~\onlinecite{cambria2023temperature} for details) and $n(\omega)$ is the occupation function for the given phonon frequency $\omega$ which depends on temperature via $n=(e^{\hbar\omega/k_B T}-1)^{-1}$. Our paper is motivated by room temperature qubit operation. Thus, we limited the acoustic phonon modes up to 40~meV in the calculation of the spin-phonon matrix elements as the higher frequency modes only contribute to the spin-phonon relaxation rates at much higher temperatures than room temperature~\cite{mondal2023spin}.\\

%\section*{Results}\label{sec:results}
%\subsection*{Monolayer boron nitride}

\begin{figure}
%    \centering
\includegraphics[width=0.45\textwidth]{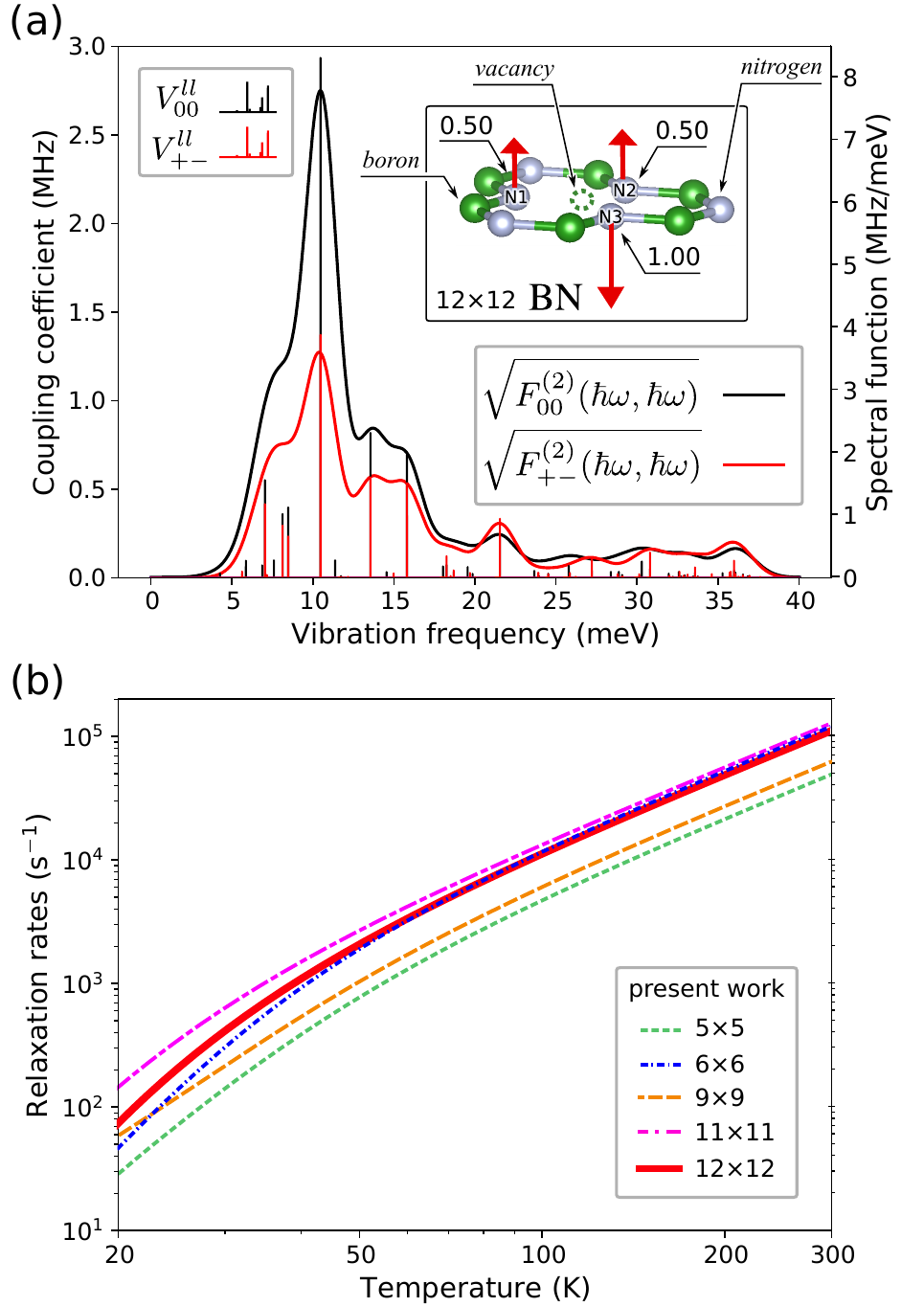}
\caption{\label{fig:spectralfBN_T1}Computed spin properties of V$_{\text{B}}^{-}$ as obtained in supercell model of freestanding monolayer BN.
    (a) The second-order spin-phonon coupling coefficients (lines) and the spectral functions (curves) for the $12\times12$ supercell size. Double-quantum transition (red) and spin-lattice dephasing (black). The out-of-plane phonon mode corresponds to the most intense spin-phonon coupling where the relative amplitude of the vibrating nitrogen atom is shown normalized to the unity (inset). (b) Temperature dependent spin-phonon relaxation rates with various supercell sizes.}
\end{figure}

\textit{Results.}
(a) \textit{Monolayer boron nitride.}
We started the investigation with the freestanding monolayer BN. We computed the spin-phonon related spectral function of  V$_{\text{B}}^{-}$ embedded in various sizes of supercells, $5\times5$ to $12\times12$. The computed spectral function for the largest considered supercell size is plotted in Fig.~\ref{fig:spectralfBN_T1}(a), while the derived spin-phonon relaxation rates for all the considered supercell sizes are depicted in Fig.~\ref{fig:spectralfBN_T1}(b). 

The spin-phonon spectral function in Fig.~\ref{fig:spectralfBN_T1}(a) shows up the most intense peak at around 11~meV.
The analysis of the phonon modes implies that a long-wavelength phonon with $\sim 11$~meV energy is an out-of-plane vibration localized at the defect site. The vast majority of the spin-phonon relaxation effectively originates from this phonon mode, which locally distorts the defect structure and thereby the spin density distribution. This analysis reinforces the phenomenological model in a recent study~\cite{liu2025temperature} where they attributed the observed spin-phonon relaxation rate to these quasilocal vibrations.

As shown in Fig.~\ref{fig:spectralfBN_T1}, V$_{\text{B}}^{-}$ in monolayer BN exhibits solely double-quantum transition between the spin states $|m_s = -1\rangle$ and $|m_s = +1\rangle$ (red curve) whereas the single-quantum transition between $|m_s = 0\rangle$ and $|m_s = \pm 1\rangle$ is forbidden due to the selection rules of the $D_{3h}$ symmetry that was previously discussed in Ref.~\onlinecite{mondal2023spin}. The similarity of the spectral function lineshapes for the spin-lattice dephasing (black curve) and the spin-lattice relaxation (red curve) indicates that these processes have similar temperature scaling. This result underscores the ultimate spin-phonon related limit for the coherence times.  

Our calculations revealed a practical issue in the \textit{ab initio} modeling of spin defects in the honeycomb lattice of BN.
For $T>50$~K, the $11\times11$ curve is very similar to the $12\times12$ curve in the computed spin-phonon relaxation rates, so the $12\times12$ supercell size seems numerically convergent for elevated temperatures. We find that the $6\times6$ curve closely follows the $12\times12$ curve in the temperature dependent spin-phonon relaxation rates because the respective spin-phonon spectral functions are similar to each other (see Appendix C). This allows us to compute these properties in bulk hBN and rBN, composed from the appropriate stacking of $6\times6$ sheets.

\begin{figure}
%    \centering
\includegraphics[width=0.45\textwidth]{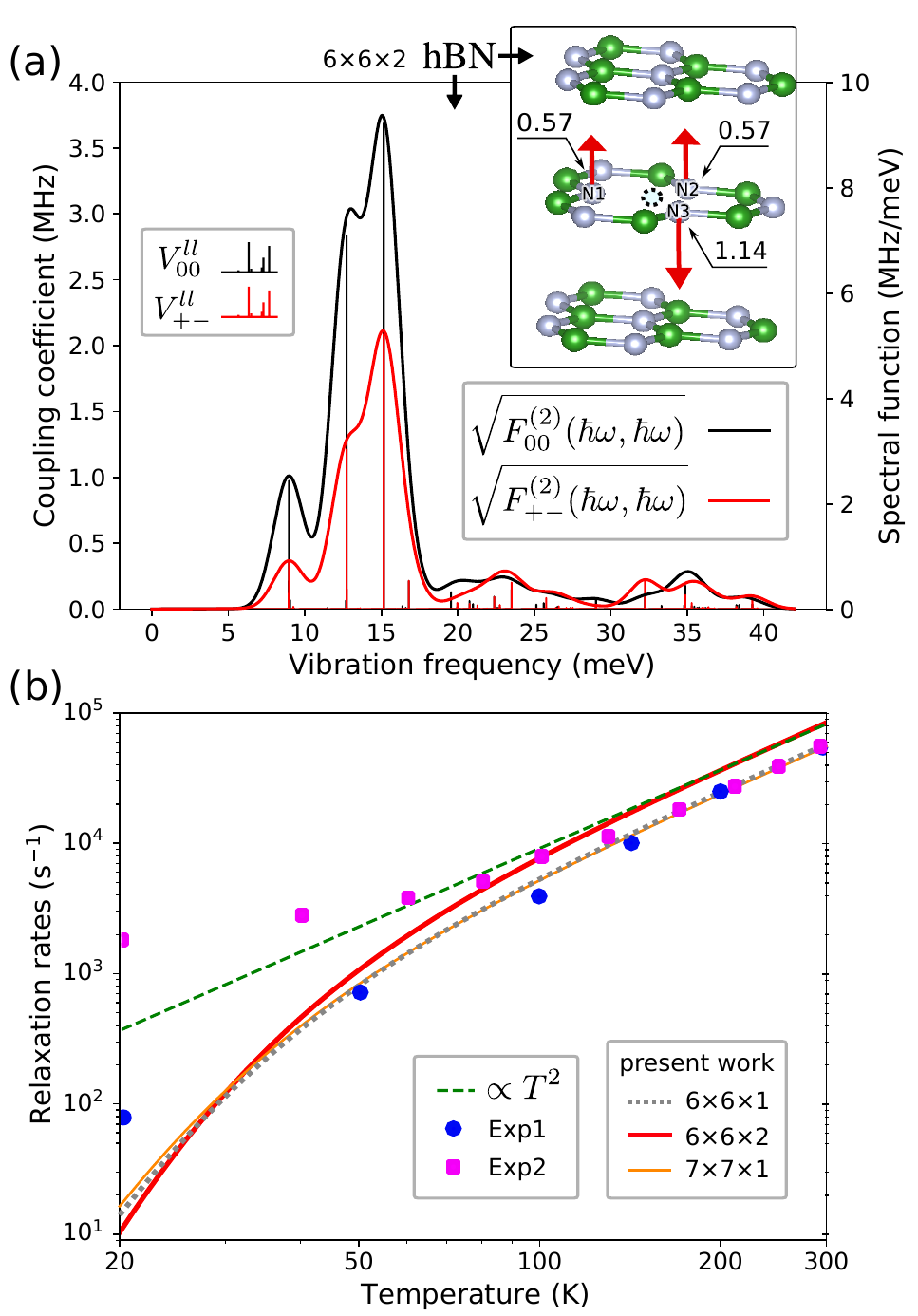}    \caption{\label{fig:spectralfhBN_T1} Computed spin properties of V$_{\text{B}}^{-}$ as obtained in $6\times6\times2$ supercell model of hBN.
    (a) The second-order spin-phonon coupling coefficients (lines) and the spectral function (curves). Double-quantum transition (red) and spin-lattice dephasing (black). The out-of-plane phonon mode corresponds to the most intense spin-phonon coupling where the relative amplitude of the vibrating nitrogen atom is shown as scaled with that in monolayer BN (inset). (b) Spin-phonon relaxation rates with various supercell sizes compared to experimental data from Exp1 (circles) (see Ref.~\onlinecite{gottscholl2021room}) and Exp2 (squares) (see Ref.~\onlinecite{liu2025temperature}). }
\end{figure}

%\subsection*{Hexagonal boron nitride}
(b) \textit{Hexagonal boron nitride.}
The analysis of phonon modes as obtained in $6\times6\times2$ supercell model of hBN shows that a phonon mode with $\sim 15$~meV energy is an out-of-plane phonon mode localized at the defect site of V$_{\text{B}}^{-}$.
The spin-phonon spectral function depicted in Fig.~\ref{fig:spectralfhBN_T1}(a) consistently shows the most intense peak at around 15~meV. It is important to note that this frequency is higher than that of the respective quasilocal vibration mode in freestanding monolayer BN. This example shows that the layer-layer interaction in realistic 2D materials can result in a quantitative difference with respect to the simple monolayer model. Apart from this important quantitative difference, the calculated spin-phonon spectral functions for V$_{\text{B}}^{-}$ spins are similar for the monolayer BN and hBN, e.g., the same type of interactions take place, the double quantum jumps for the spin-phonon relaxation. This is expected because V$_{\text{B}}^{-}$ defects in monolayer BN and hBN share the same $D_{3h}$ symmetry~\cite{mondal2023spin}, thus they share the same selection rules.

The computed spin-phonon relaxation rates of V$_{\text{B}}^{-}$ compared to existing experimental data in hBN are depicted in Fig.~\ref{fig:spectralfhBN_T1}(b). The two experimental data deviate for $T<100$~K. This issue implies that the observed $T_1$ times in the two experiments~\cite{gottscholl2021room, liu2025temperature} for low temperatures are not solely governed by spin-phonon relaxation. For $T>100$~K, both the experimental data and the computed values show the same trend which yields the $T^2$ power-law. The calculated rate at room temperature ($\sim 7\times10^4$~Hz) is very close to the experimental one ($\sim 5\times10^4$~Hz) which corresponds to 12.5~$\mu$s and 18~$\mu$s $T_1$ times, respectively. 
%
% Origin of the $T^2$ power-law?
%

By approaching cryogenic temperatures, the computed spin-phonon relaxation rates become much slower than the experimental ones. As anticipated above, the direct comparison of the experimental data and the computed values are not straightforward because other factors than spin-phonon relaxation may enter. Nevertheless, the calculated spin-phonon relaxation rate is likely underestimated because of the lack of low-energy phonon modes in the finite supercell size model that may couple to the spin. On the other hand, the spin-phonon relaxation rates can be well computed for the $100<T<300$~K region with our method.

%\subsection*{Rhombohedral boron nitride}
(c) \textit{Rhombohedral boron nitride.}
We continue the investigation of the spin-phonon relaxation in 2D materials with the case of V$_{\text{B}}^{-}$ in rBN. In rBN, the ABC stacking of honeycomb BN lattices will result a lower symmetry group, $C_{3v}$, for the V$_{\text{B}}^{-}$ defect. Furthermore, the layer-layer interaction should differ in hBN and rBN due to the different stacking in the two materials.

\begin{figure}
%    \centering
\includegraphics[width=0.45\textwidth]{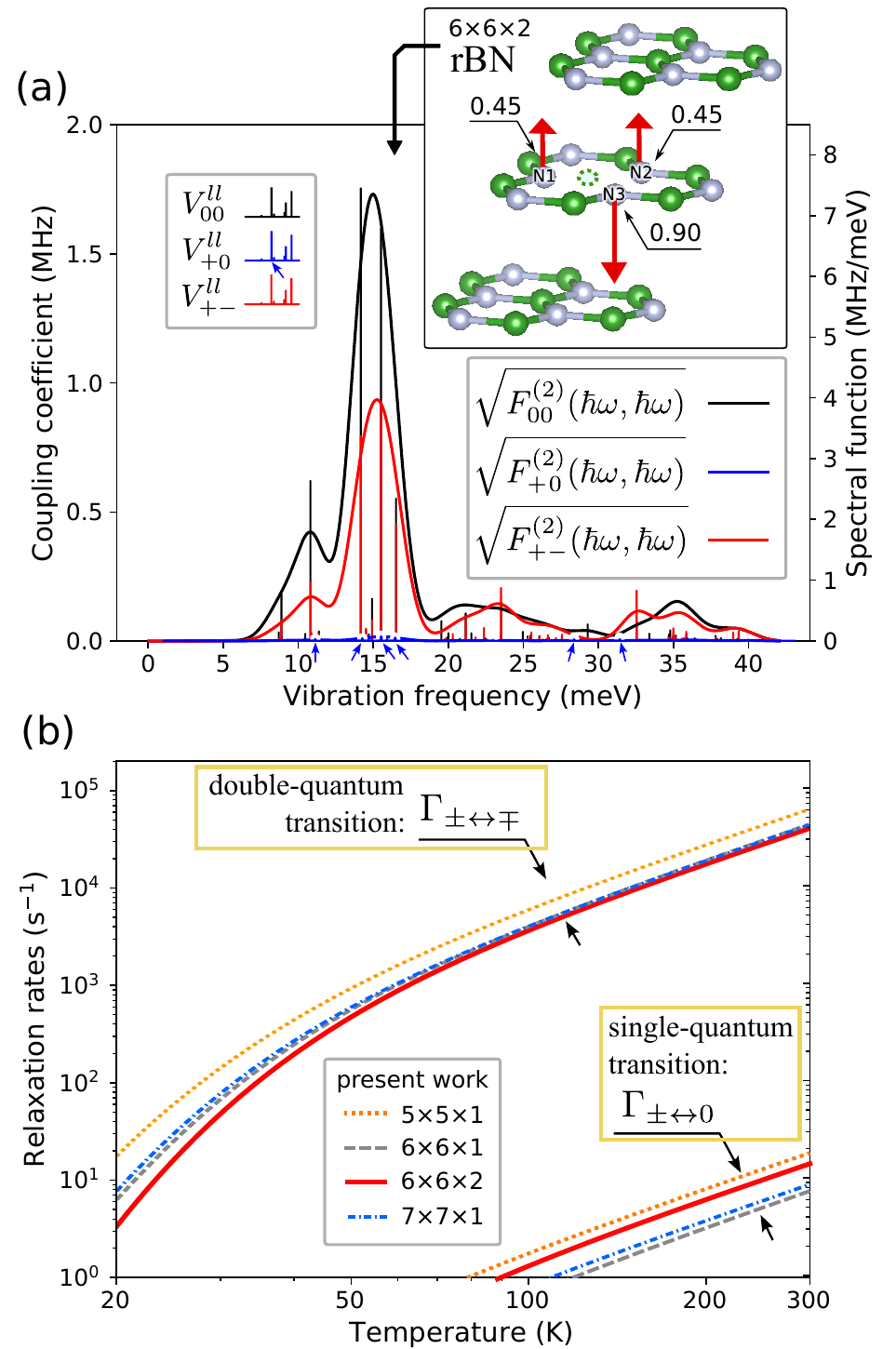}
\caption{\label{fig:spectralfrBN_T1} Computed spin properties of V$_{\text{B}}^{-}$ as obtained in $6\times6\times2$ supercell model of rBN. (a) Spin-phonon spectral function with double-quantum transition (red), single-quantum transition (blue), and spin-lattice dephasing (black). The out-of-plane phonon mode corresponds to the most intense spin-phonon coupling where the relative amplitude of the vibrating nitrogen atom is shown as scaled with that in monolayer BN (inset). (b) Spin-phonon relaxation rate with double-flip transition with various supercell sizes.}
\end{figure}

The analysis of phonon modes in the $6\times6\times2$ supercell model of rBN shows that the phonon mode with energy $\sim15.5$~meV corresponds to out-of-plane vibration. This phonon mode appears close to the most intense peak position of the spin-phonon spectral function [see Fig. \ref{fig:spectralfrBN_T1}(a)].

The reduced symmetry in rBN opens the single quantum transition channel for spin-phonon relaxation that can principally accelerate the spin-phonon relaxation rates. This channel is plotted as a blue curve in Fig. \ref{fig:spectralfrBN_T1}(b). As can be seen its contribution to the spin-phonon spectral function is minor and the predominant spin-phonon relaxation path still goes with double quantum transition [c.f., red and blue curves in Fig.~\ref{fig:spectralfrBN_T1}(a)].

%
%This result is in agreement with the low-energy out-of-plane phonon at $\sim 20$~meV, which is directly observable in experiments as an asymmetric broadening of ZPL \cite{iwanski2024revealing}. 
%
%

%
%\noindent \textit{Discussion} Compared to the NV center in diamond, the ground state ZFS of $V_{B}^{-}$ in hBN with $AA'$ stacking exhibits a $\sim$ 25 times larger variation from cryogenic to room temperatures, placing it in an advantageous position for nanoscale thermometry \cite{liu2025temperature}.
%This property is due to a much higher activation temperature of NV compared to $V_{B}^{-}$. 

%\section*{Discussion} \label{sec:discussion}
\textit{Discussion.}
By comparing the spin-phonon relaxation rates of V$_{\text{B}}^{-}$ in the considered 2D BN materials as plotted in Fig.~\ref{fig:structs_rates}(b), one finds an universal $T^2$ scale power-law in $150<T<300$~K region. The calculated spin-phonon spectral functions of these systems exhibit common features: the spectral functions may be approximated by a single effective phonon that couples to the spin and there is a relatively abrupt cutoff phonon frequency where the spin-phonon coupling strongly reduces. In this scenario, the spin relaxation rate may be simplified to
\begin{equation}
\label{eq:gamma}
\Gamma(T)= 1/T_1=\sum_{i}{A_i}{n_i(\omega)}(n_i(\omega)+1)+A_s
\end{equation}
where $A_i$ are the coupling coefficients associated with the effective modes, and $A_s$ is a sample-related constant. When $k_\text{B}T\gg \hbar \omega_0$ where $k_\text{B}$ is the Boltzmann constant and $\omega_0$ is the cutoff phonon frequency then $n_i \propto T$ so the leading term in $n_i(\omega)(n_i(\omega)+1)$ also goes with $T^2$ which basically agrees with a recent derivation~\cite{tabesh2025decoherence}. One might conclude that this universal scale is not directly related to the 2D nature of the host material as the cutoff frequency in the spin-phonon spectral function can principally occur in 3D systems too. For the specific defect in our study, out-of-plane quasilocal vibration modes are mostly coupled to the defect spin that have low frequencies. In this sense, the universal $T^2$ power-law can be associated with the 2D nature of the host material in which the out-of-plane phonon modes are specific to 2D materials. We expect similar behavior for other planar point defect $S=1$ spins in 2D materials. 

Although the temperature-dependent spin-phonon relaxation rates exhibit similar slopes, noticeable shifts in the curves can be observed in Fig.~\ref{fig:structs_rates}(b), despite the identical BN host layers and the same type of defect spins—features not captured by phenomenological models. Comparing monolayer BN and hBN, we find that the computed spin-phonon relaxation rates of V$_{\text{B}}^{-}$ are nearly identical in the temperature range $100 < T < 300$~K. In contrast, the relaxation rates in rBN are generally slower than those in hBN. The spin-phonon relaxation rate at a given temperature depends on the effective phonon frequency ($\omega$) through the phonon occupation number $n(\omega)$ and the strength of the spin-phonon coupling at that frequency. For instance, a blueshift in $\omega$ or a reduction in coupling strength would result in a slower relaxation rate. By comparing the spin-phonon spectral functions of V$_{\text{B}}^{-}$ in monolayer BN and hBN (see Figs.\ref{fig:spectralfBN_T1} and \ref{fig:spectralfhBN_T1}), we observe two competing trends: (i) the effective phonon frequencies shift to higher energies in hBN, and (ii) the spin-phonon coupling is generally stronger in hBN. These effects nearly cancel each other, resulting in very similar relaxation rates in the $100 < T < 300$~K range, although the rates in hBN are slightly slower. This explains the success of the simplified monolayer BN model~\cite{mondal2023spin} in closely reproducing the experimental spin-relaxation rates observed for V$_{\text{B}}^{-}$ in hBN~\cite{gottscholl2021room}.

We find that nitrogen atoms with dangling bonds exhibit larger vibrational amplitudes in hBN than in rBN or monolayer BN [see insets of Figs.~\ref{fig:spectralfBN_T1}(a), \ref{fig:spectralfhBN_T1}(a), and \ref{fig:spectralfrBN_T1}(a)]. These enhanced vibrations in hBN more strongly perturb the spin density matrix, thereby increasing the spin-phonon coupling. This effect likely originates from the attractive electrostatic interaction between positively polarized nitrogen atoms and negatively polarized boron atoms, which are aligned directly above one another in the AA$^\prime$ stacking configuration---a distinctive structural feature of hBN. We propose that this is the microscopic origin of the variation in spin-phonon coupling strengths across the 2D BN polytypes. Notably, the spin-phonon coupling strength in rBN is approximately four times weaker than in hBN (see Figs.\ref{fig:spectralfhBN_T1} and \ref{fig:spectralfrBN_T1}), while the effective phonon frequencies remain comparable. As a result, the spin-lattice relaxation of V$_{\text{B}}^{-}$ is slower in rBN, as confirmed in Fig.~\ref{fig:structs_rates}(b), although the corresponding $T_1$ times are of the same order of magnitude. Nevertheless, these findings suggest that V$_{\text{B}}^{-}$ exhibits superior spin coherence properties in rBN compared to hBN.

Our results also provide insight into the interpretation of recent observations regarding the temperature dependence of $T_1$ for V$_\text{B}^-$ centers in ultrathin hBN flakes~\cite{Durand2023}. Compared to bulk hBN, the $T_1$ time measured at room temperature in ultrathin flakes was significantly shorter (approximately 1~$\mu$s), a phenomenon previously attributed to fluctuating magnetic fields generated by surface paramagnetic impurities. Notably, this $T_1$ time increased by three orders of magnitude at cryogenic temperatures. Our findings suggest that thermally activated magnetic noise---such as phonon-mediated fluctuations of surface paramagnetic impurities---could explain this temperature dependence, rather than invoking a modified spin-phonon interaction in few-layer hBN.

%\section*{Conclusions}%
\textit{Conclusions.}
In conclusion, the temperature dependence of spin-phonon relaxation for the V$_{\text{B}}^{-}$ defect spin was investigated in various BN lattices to understand the role of interlayer interactions in 2D materials during the relaxation process. While the spin-phonon relaxation rates of V$_{\text{B}}^{-}$ scale with a common slope as a function of temperature, discernible shifts in the curves are observed among monolayer BN, hBN, and rBN. These differences are attributed to significant variations in the frequency and amplitude of out-of-plane quasilocal vibrational modes that couple to the defect’s electron spin. This behavior arises from interlayer interactions, highlighting that phenomenological theories alone cannot capture such subtle---but important---differences in 2D systems. Instead, accurate \textit{ab initio} computations are required to reveal the underlying physical mechanisms. In the specific case of the V$_{\text{B}}^{-}$ defect, we find that it exhibits superior spin properties in rBN compared to hBN. This enhancement may benefit practical applications in quantum technologies, such as nanoscale sensing of temperature and magnetic fields.

\textit{Acknowledgments.} 
The authors thank Mehdi Abdi for
his comments. We gratefully acknowledge the support of
the Quantum Information National Laboratory of Hungary,
funded by the National Research, Development, and Innovation Office of Hungary (NKFIH) under Grant No.\
2022-2.1.1-NL-2022-00004. G.T.\ was supported by the J\'anos
Bolyai Research Scholarship of the Hungarian Academy of
Sciences and by NKFIH under Grant No.\ STARTING 150113.
A.G.\ acknowledges access to high-performance computational resources provided by KIFÜ (Governmental Agency
for IT Development, Hungary) and funding from the European Commission for the QuMicro (Grant No.\ 101046911)
and SPINUS (Grant No.\ 101135699) projects, as well as the
QuantERA II project Maestro (NKFIH Grant No.\ 2019-2.1.7-ERA-NET-2022-00045). N.E.\ acknowledges the scholarship
support from the Iranian Ministry of Science, Research, and
Technology. This work is based upon research funded by the
Iranian National Science Foundation (INSF) under Project
No.\ 4021945.

A.G.\ devised and planned all the calculations, while N.E.\
performed the calculations. I.A.S., G.T., and A.G.\ supervised
N.E., and contributed to the discussion of the results. N.E.\ and
A.G.\ wrote the initial draft of the paper, incorporating input
from all co-authors. A.G.\ supervised the project.

The authors declare no competing interests.

%\subsection*{Data availability}
\textit{Data availability.} The data that support the findings of this
article are not publicly available. The data are available from
the authors upon reasonable request.

%\appendix
%\section{Zero-field splitting tensor data}
\textit{Appendix A: Zero-field splitting tensor data.}
In the global energy minimum of the adiabatic potential energy surface, the $D$-constant is equal to $(3/2)D_{zz}$ after diagonalization of the $D$-tensor.
Here we do not apply spin decontamination procedure for saving computational time~\cite{montoya2000spin, biktagirov2020spin}. We assume that the same error occurs in the $D$-tensor for the undistorted and distorted structures which results in accurate derivatives. We used this assumption for computing the spin-phonon relaxation time of nitrogen-vacancy center in diamond~\cite{cambria2023temperature} which worked well for that system. The computed PBE $D$-tensors with various supercell sizes are given in Table~\ref{tab:convergent_size_of_ZFS}. As can be seen, the absolute $D$-constant values are close to each other. This can be understood from the largely localized nitrogen dangling bonds that builds up the spin density matrix.

We note that applying spin decontamination procedure the calculated $D$-constant agrees well with the observed one (3.5 GHz) in hBN~\cite{gottscholl2020initialization, ivady2020ab}. The computed $D$-constants with Heyd-Scuseria-Ernzerhof (HSE) functional~\cite{heyd2003hybrid} with $\alpha=0.32$ parameter together with spin-decontamination procedure are $3.19$~GHZ and $3.44$~GHz in $9\times9\times1$ hBN and rBN supercells, respectively. The latter is close to a recently reported value of $3.45$~GHz for an $S=1$ defect spin observed by optically detected magnetic resonance in irradiated rBN samples~\cite{gale2025quantum}.

\begin{table*}
%    \centering
 %    \renewcommand{\arraystretch}{1.3}
\caption{\label{tab:convergent_size_of_ZFS}Convergence test of ZFS with supercell size.}
 %   \begin{tabular}{|c @{\quad} c| c @{\quad} c| c @{\quad} c|}
 \begin{ruledtabular}
 \begin{tabular}{cccccc}
 %       \hline
         Monolayer & ZFS (GHz) & 
         hBN & ZFS (GHz) & 
         rBN & ZFS (GHz) \\
        \hline
        $5\times5\times1$ & 2.92 &  $5\times5\times1$ & 2.64 &  $5\times5\times1$ & 2.83 \\
        $6\times6\times1$ & 2.83 &  $6\times6\times1$ & 2.56 &  $6\times6\times1$ & 2.77 \\
        $7\times7\times1$ & 2.79 &  $7\times7\times1$ & 2.52 &  $7\times7\times1$ & 2.73 \\
        $9\times9\times1$ & 2.75 &  $9\times9\times1$ & 2.48 &  $9\times9\times1$ & 2.70 \\
        $11\times11\times1$ & 2.74 & $6\times6\times2$ & 2.84 & $6\times6\times2$ & 2.88 \\
        $12\times12\times1$ & 2.74 &  $6\times6\times3$ & 2.89 & $6\times6\times3$ & 2.91 \\
%        \hline
    \end{tabular}
 \end{ruledtabular}  
\end{table*}

%\section{Calculation method of the phonons and phonon density of states}
\textit{Appendix B: Calculation method of the phonons and phonon density of states}
Atoms are moved by 0.01~\AA\ in each direction to calculate the total energy by DFT-PBE, and the resulting adiabatic potential energy surface is fit to a parabola around the global energy minimum to build up the Hessian matrix for calculating phonon eigenfunctions (normal coordinates) and eigenvalues (frequencies).

\begin{figure*}
%    \centering
\includegraphics[width=1\textwidth]{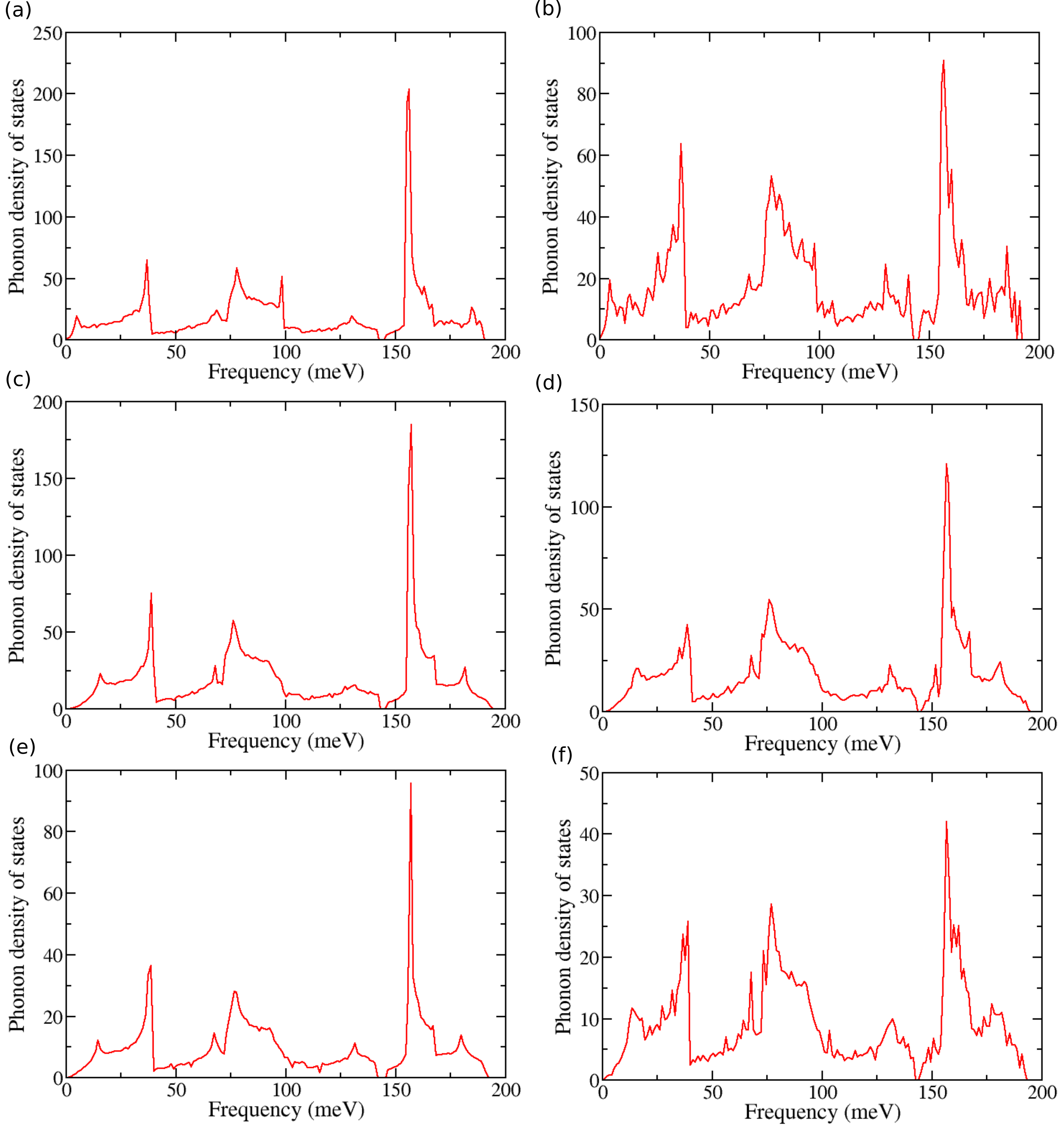}
    \caption{\label{fig:phononDOS} 
    Computed phonon density of states for pristine and defective supercells with V$_{\text{B}}^{-}$. 
    (a) Pristine $12\times12$ monolayer BN, (b) defective $12\times12$ monolayer BN, (c) pristine $6\times6\times2$ hBN, (d) defective $6\times6\times2$ hBN, (e) pristine $6\times6\times2$ rBN, (f) defective $6\times6\times2$ rBN.}
\end{figure*}

The phonon density of states (DOS) for various BN systems are plotted in Fig.~\ref{fig:phononDOS}. We find that the pristine [Fig.~\ref{fig:phononDOS}(a)] and defective [Fig.~\ref{fig:phononDOS}(b)] supercell phonon DOS for the freestanding monolayer are very similar which is expected for a vacancy defect. We plot the phonon DOS for the defective hBN and rBN in Fig.~\ref{fig:phononDOS}(d) and Fig.~\ref{fig:phononDOS}(f), respectively.  As can be seen, there is an intense peak at around 5-15~meV in the phonon DOS that experiences a blue shift going from monolayer BN to rBN. We associate most of the phonons around this region with the out-of-plane motion of ions. In the 15-40~meV region another peak is developed at around 40~meV. However, those phonons are associated with the motion of ions within the sheet that do not couple to the defect spin. Therefore, the computed spin-phonon spectral function will be relatively small in that region.

\begin{figure*}
%    \centering
\includegraphics[width=1\textwidth]{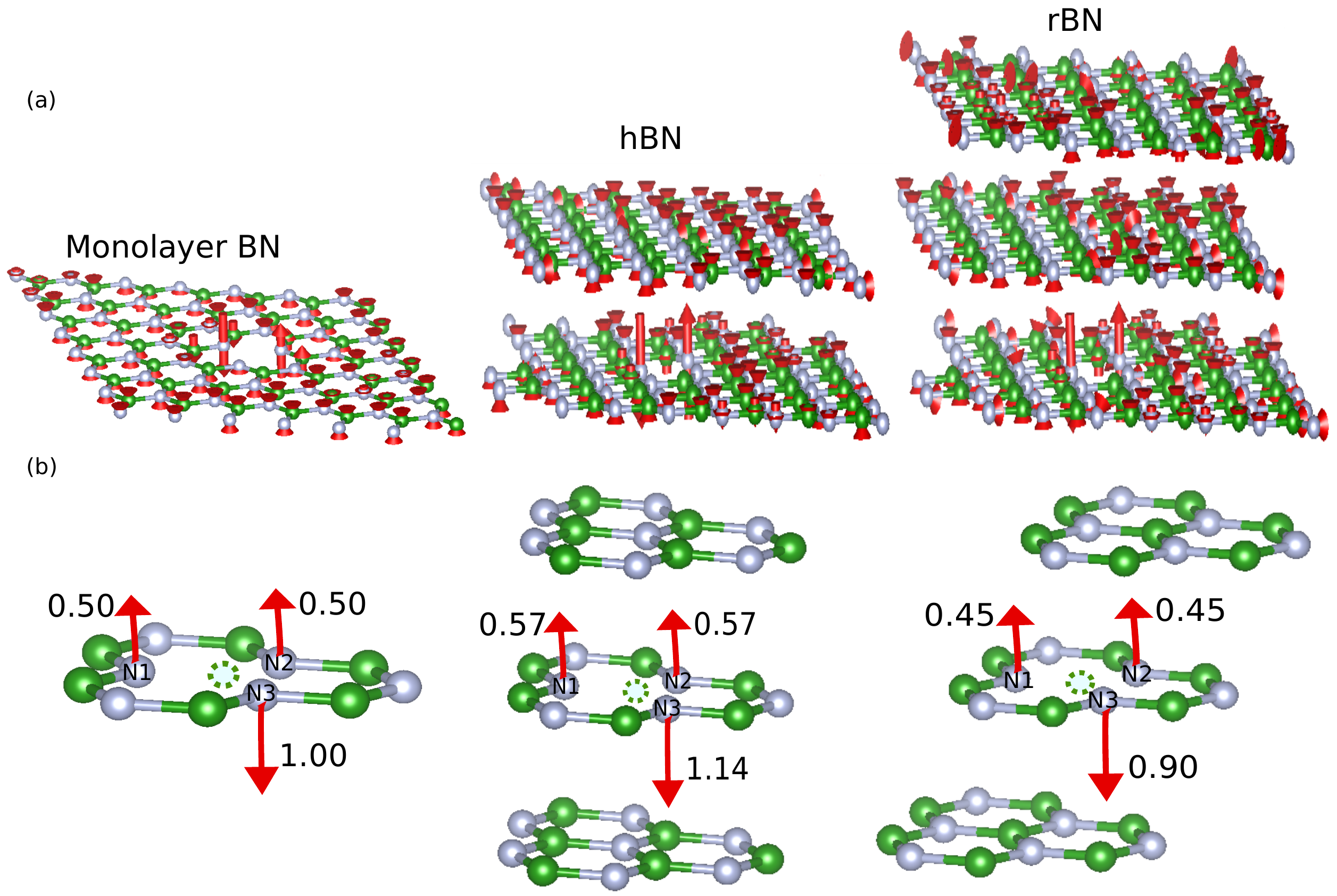}
    \caption{\label{fig:exph} 
    Quasilocalized $e_x$ phonon of V$_{\text{B}}^{-}$ is plotted for monolayer BN, hBN, and rBN. (a) Motion of ions in the supercells that are represented by arrows. (b) Zoomed in structure around the vacancy defect where the relative amplitude of the vibrating nitrogen atom is shown normalized to the unity in the monolayer BN and the relative amplitudes are shown in hBN and rBN.
    }
\end{figure*}

We also plot the motion of ions associated with the $e_x$ phonon that couples to the defect's spin most intensely (Fig.~\ref{fig:exph}). In the main text, we zoom in to the central part of the defect where we show the motion of nitrogen atoms near the vacancy with the relative amplitudes.

\begin{figure*}
%    \centering
\includegraphics[width=1\textwidth]{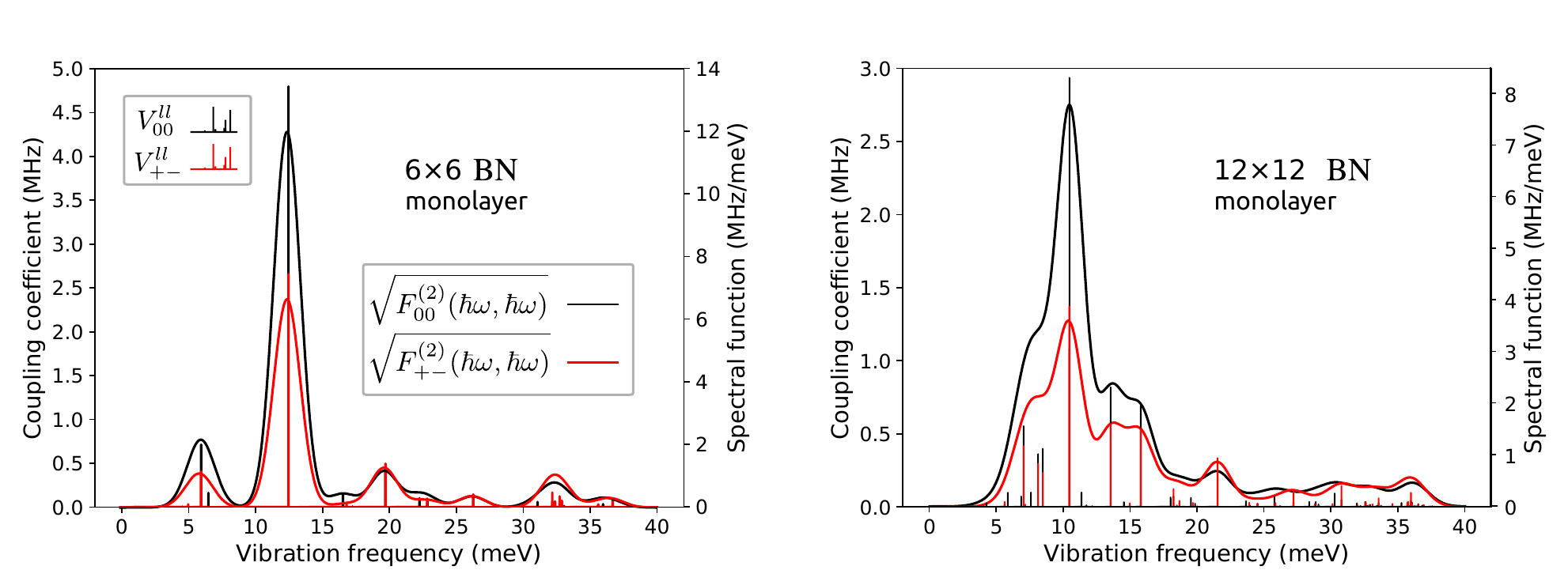}
\caption{\label{fig:spf66_vs_1212} 
    Spin-phonon spectral functions of V$_{\text{B}}^{-}$ as obtained in $6\times6$ and $12\times12$ supercell model of freestanding monolayer BN.
    The second-order spin-phonon coupling coefficients (lines) and the spectral functions (curves) are plotted. Double-quantum transition (red) and spin-lattice dephasing (black). The out-of-plane phonon mode corresponds to the most intense spin-phonon coupling (inset, central part of structure).}
\end{figure*}

%\appendix
%\section{Spin-phonon spectral function: additional data}
\textit{Appendix C: Spin-phonon spectral function: additional data.}
For the calculation of the second-order spin-phonon spectral fuction, all degenerate $e_x$, $e_y$ phonon modes of the supercell distort the trigonal symmetry of the system while the symmetric $a_1$ phonon modes keep the trigonal symmetry of the defect intact. To calculate the derivatives along these normal coordinates we applied the step of displacement $\sqrt{(\Delta R)^2}= 0.1$~\AA $\sqrt{\text{a.m.u.}}$ along the respective normal coordinates.
We find that the calculated spin-phonon spectral functions of V$_{\text{B}}^{-}$ as obtained in $6\times6$ and $12\times12$ supercell are similar to each other as plotted in Fig.~\ref{fig:spf66_vs_1212}.

%\section{References}
%\noindent \textbf{References}

\bibliography{references}% Produces the bibliography via BibTeX.

\end{document}